\documentclass[twocolumn]{aastex701}

\usepackage{romannum}
\usepackage{CJKutf8}
\usepackage{booktabs}
\usepackage{multirow}
\usepackage{amsmath}
\usepackage{float}
\usepackage{appendix}
\usepackage{ragged2e}
\usepackage{xcolor}
\usepackage{soul}
\soulregister\ref7
\soulregister\Romannum7

\newcommand{\CJKtext}[1]{\begin{CJK*}{UTF8}{bsmi}(#1)\end{CJK*}}

\newcommand{\hd}{$\mathrm{EW(H\delta_A)}$ }
\renewcommand{\thefootnote}{\fnsymbol{footnote}}
\begin{document}

\title{Exploring the Origin of Rejuvenating Gas from MaNGA Nearby Galaxies}

\author[0009-0002-9423-2373]{Ting-Hsuan Li \CJKtext{李廷軒}}
\affiliation{Graduate Institute of Astrophysics, National Taiwan University, Taipei City 106216, Taiwan}
\email[]{r11244005@ntu.edu.tw}

\author[0000-0002-9665-0440]{Po-Feng Wu \CJKtext{吳柏鋒}}
\affiliation{Graduate Institute of Astrophysics, National Taiwan University, Taipei City 106216, Taiwan}
\email[]{wupofeng@phys.ntu.edu.com}


\begin{abstract}

This study investigates the origin of secondary star formation, i.e., rejuvenation in nearby galaxies. From the MaNGA IFU survey, we use stellar absorption features, D$_n$4000 and EW(H$\mathrm{\delta_A}$), to identify regions that started the rejuvenation within the last $\sim$200~Myr and use gas-phase metallicity as a primary tracer for accretion of pristine gas, in order to verify the mechanism that triggers the rejuvenation. 
We compare the metallicity and the velocity of rejuvenating regions with typical star-forming regions. We also compare metallicity gradients, environments, \ion{H}{1} gas fractions, and visual morphologies of galaxies hosting rejuvenating regions to controlled star-forming and quiescent galaxy samples. Overall, we do not find the rejuvenating regions or their hosts show anomalies in metallicity, kinematics, and visual morphology to the controlled comparison samples. These observations suggest that local rejuvenation is likely fueled by gas already residing within the host rather than accreted from outside, and reflect the short life time of small scale dense gas clouds. On the contrary, if the rejuvenation is fueled by gas accretion, the chemical and mixing timescale should be much shorter than $\sim100$ Myr so that no chemical and kinematical anomalies are measured. 
Meanwhile, we report a clear case of a massive quiescent galaxy brought back to star-forming by accreting gas. Our method of identifying rejuvenation is simple and effective, and can be applied to large spectroscopic surveys to investigate the origin of rejuvenation across cosmic time.
\end{abstract}


\section{Introduction} \label{sec:intro}
Galaxies exhibit a bimodal distribution on color-color diagrams \citep{2001Strateva, 2004Baldry}, where blue galaxies are actively forming stars and quiescent galaxies have insignificant growth of in-situ star formation activities. Generally, galaxies evolve from being star-forming to quiescent, suggested by the increasing number density of quiescent galaxies with cosmic time \citep{2013Ilbert,2013Muzzin}. However, not all galaxies follow this general trend at all time. Various observations show that some galaxies can have a secondary star formation after they once became quiescent \citep{2007Kaviraj,2007Donas,2007Schawinski,2010Salim,2010Thilker,2012Fang,2019Chauke,2022Rathore,2024Tanaka}. Such revival of star formation activity is often referred as ``rejuvenation''.

These rejuvenation galaxies indicate that the quiescent status can be a temporary state. At $z\simeq0$, both cosmological hydrodynamical simulation and constraints from the stellar continua of galaxies suggest that $\sim10\%$ of galaxies had experienced at least one rejuvenation \citep{2018Nelson,2024Tanaka}. The unsolved question is: what is the origin of the gas that fuels this renewed star formation, and what physical processes trigger its collapse into new stars? On one hand, the star-forming gas could be obtained through accretion of gas from the circumgalactic or intergalactic medium or via mergers with smaller, gas-rich galaxies \citep{2009Kaviraj,2015Mapelli,2018Diaz,2020Werle,2022Rathore}. On the other, rejuvenation can be fueled by pre-existing reservoir of gas within the galaxy that was not fully consumed \citep{2012Thom,2022Woodrum,2023Paspaliaris}.

The gas-phase metallicity provides a powerful diagnostic to distinguish the origin of gas. Gas accreted from external sources is expected to be relatively metal-poor and should lead to a noticeable dilution of the gas-phase metallicity in the star-forming regions \citep{2016Ceverino,2019Hwang,2025Estrada-Carpenter}. The rejuvenating gas may also exhibit distinct kinematics from the galaxy \citep{2013Bouche, 2016Bouche,2019Ho,2024Coleman,2026Bevacqua}. For example, observationally, kinematically distinct inflowing gas has been detected via redshifted ISM absorption lines, with inflowing velocities of $\sim 250-280$ km/s relative to the host galaxy's systemic velocity \citep{2024Coleman,2026Bevacqua}; according to simulations, gas accreted onto galaxies typically has inflow speeds of $\sim 20-60$ km/s \citep{2019Ho}. On the contrary, if pre-existing gas is the fuel, the gas metallicity in the rejuvenation regions would not deviate much from the typical values. Furthermore, gravitational interaction with another galaxy can drive inflow of gas from the outskirts to center. The gas flow can elevate star formation and flatten the metallicity gradient at the same time \citep{2010Rupke_a,2018Bustamante,2019Moreno}. Probing the metallicity gradients of galaxies could serve a way to identify recent gas flows in galaxies. 

In this work, we attempt to identify galaxies that are currently experiencing rejuvenation in the Mapping Nearby Galaxies at APO \citep[MaNGA;][]{Bundy2015} integral field unit (IFU) survey. The properties of these galaxies provide the best proxies to understand the possible mechanisms that drive rejuvenation. MaNGA provides precise measurements of stellar absorption features of galaxies, allowing us to construct a clean sample of rejuvenation events in galaxies. We can then estimate gas metallicities of each rejuvenation region, as well as the metallicity gradient of each host galaxy to investigate the origin of the rejuvenating gas. In addition, using the profound information in the IFU data, we will demonstrate that, as suggested by stellar population synthesis models, current rejuvenation events can be identified from only 2 strongest stellar absorption features, D$_n$4000 and \hd \citep{2023Zhang}. 

This paper is constructed as follows. We describe our data and sample selection in Section \ref{sec:data}. We present the properties of galaxies hosting rejuvenation regions in Section \ref{sec:local_RJGs}. In Section~\ref{sec:special}, we show an unambiguous case that a rejuvenation event triggered by accretion is caught in the act. We then discuss the implication from our results in Section \ref{sec:discussion} and summarize this work in Section~\ref{sec:conclusion}. Throughout this work, we assume $H_0=70 \text{ km s}^{-1}\text{Mpc}^{-1}$, $\Omega_m=0.3$, and $\Omega_\Lambda=0.7$.

\section{Data and Sample} \label{sec:data}

\subsection{The MaNGA Survey}
MaNGA \citep[Mapping Nearby Galaxies at APO;][]{Bundy2015} is a part of the fourth-generation Sloan Digital Sky Survey \citep[SDSS-\Romannum{4};][]{Blanton2017}, which observed approximately 10,000 nearby galaxies in the redshift range $0.01<z<0.15$ using Integral Field Units (IFUs). MaNGA observes each galaxy in the wavelength range from 3600\r{A} to 10300\r{A} at resolution $R\approx2000$. We use the data release (DR17) \citep{2022DR17} in this work.

We use spectral indices and emission line fluxes of each spaxels from MaNGA Data Analysis Pipeline \citep[DAP;][]{Belfiore2019, Westfall2019}, and the redshift, axial ratio ($b/a$), and NUV and $r$ magnitudes of galaxies from MaNGA Data Reduction Pipeline \citep[DRP;][]{Law2016}.

\subsection{The Value-Added Catalogs (VACs)}
This work also uses several MaNGA Value-Added Catalogs (VACs). We take stellar masses and stellar mass densities from \textsc{Pipe3D} VAC \citep{2016Pipe3D, 2022Pipe3D}. We correct the observed stellar mass density by $\Sigma_\ast = \Sigma_{\ast, \text{obs}}\times (b/a)$ due to the inclination of galaxy.

We take the numerical morphological types from the \textsc{MaNGA Visual Morphologies} \citep{2022Vazquez}. The visual classification is based on the $r$-band images of SDSS and DESI Legacy Survey \citep{2019Dey}.

We quantify the environments of galaxies using the \textsc{Galaxy Environment for the MaNGA value added catalog} \citep[GEMA-VAC;][]{2015Argudo}, in which defines primary galaxies with a range of \textit{r}-band model magnitude $11\le m_r \le15.7$ and neighboring galaxies with $m_r>17.7$. We use two measurements. First, the parameter $Q$ quantifies the tidal strength that the neighboring galaxies produce on the target galaxy:
\begin{equation}\label{eq:tidal_Q}
    Q \equiv \log \left[ \sum_i \frac{M_i}{M_P} \left(\frac{D_P}{d_i}\right)^3 \right],
\end{equation}
where $M_P$ and $D_P$ represent the stellar mass and diameter of the primary galaxy; $M_i$ and $d_i$ represent the stellar mass and projected physical distance of the $i$th neighbor to the primary galaxy. \citet{2015Argudo} provides two $Q$ parameters: from galaxies with radial velocities differences smaller than 500 km/s and within projected distances 1~Mpc ($Q_\mathrm{1Mpc}$) and 5~Mpc ($Q_\mathrm{5Mpc}$). Second, we use the 5th-nearest-neighbor projected density $\eta_5$ based on neighbors within 1~Mpc ($\eta_{5,\mathrm{1Mpc}}$) and 5~Mpc ($\eta_{5,\mathrm{5Mpc}}$), respectively. If a galaxy has $k$ neighbors within the specified radius and $k<5$, then $\eta_k$ is used.

We obtain the \ion{H}{1} masses of galaxies from the \textsc{\ion{H}{1}-MaNGA} DR3 \citep{2019Masters, 2021Stark}. A total of 6358 MaNGA galaxies are included in this catalog. 

\subsection{The Rejuvenating Sample}
Accurately recovering the true formation histories from observables is a challenging task, especially for galaxies with rising SFRs, where the most recent star formation event can outshine older events. Therefore, instead of performing full-spectral fittings to identify regions with current rejuvenation, we select recently rejuvenating regions using the equivalent widths of the H$\delta$ absorption, \hd \citep{1997Worthey}, and the 4000 \AA\ break, D$_n$4000 \citep{1999Balogh}. By definition, the rejuvenating regions are currently star-forming but have low SFR in the recent past. The rejuvenating regions thus have low D$_n$4000 that is comparable to normal star-forming galaxies but relatively weak \hd due to the lack of A-type stars.

\begin{figure}[htbp]
    \centering
    \includegraphics[width=\linewidth]{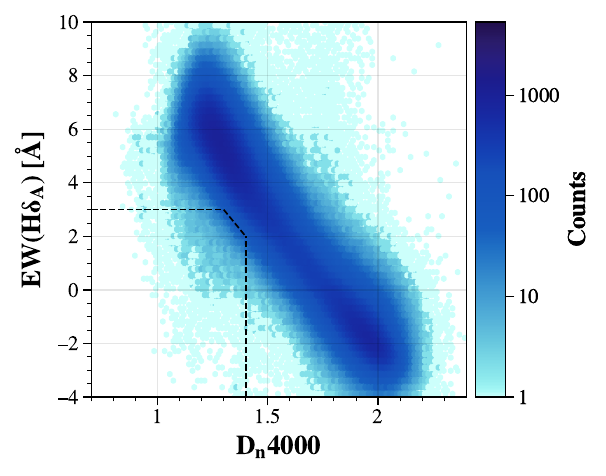}
    \caption{Distribution of all MaNGA spaxels ($N\sim 3\text{M}$) on D$_n$4000-\hd plane, visualized using hexagonal binning. The color bar on the right indicates the number of data points in each hexagonal bin. The majority of spaxels distributes tightly on a diagonal sequence, and only a small fraction fall within our selection criteria (black dashed line).}
    \label{fig:dn_hd}
\end{figure}

Figure~\ref{fig:dn_hd} shows the distribution of \hd and D$_n$4000 of MaNGA spaxels. The majority of spaxels are located on a diagonal sequence. Younger, star-forming galaxies have low D$_n$4000 and high EW(H$\delta_\mathrm{A}$), while older, quiescent galaxies occupy the high D$_n$4000 and low \hd region. The rejuvenating regions are located at the lower-left corner in Figure~\ref{fig:dn_hd}. \citet{2023Zhang} has demonstrated that combining D$_n$4000 and \hd is able to select a clean sample of current rejuvenation events. We first select spaxels that fulfill the following criteria:
\begin{subequations}\label{eq:Seletion_criteria}
\begin{gather}
    \mathrm{D}_n4000<1.4 \\
    \mathrm{EW}(\mathrm{H\delta_A}) < 3 \text{ \AA} \\
    10 \times \mathrm{D}_n4000+\mathrm{EW}(\mathrm{H\delta_A})/\text{\AA}-16 < 0, 
\end{gather}
\end{subequations}
and signal-to-noise ratios (S/N) larger than 3.

Then, we keep spaxels whose line emission is dominated by star-forming activities so that we can use emission line ratios to infer the gas metallicities (Section~\ref{sec:local_RJGs}). We use the BPT diagram \citep{1981BPT}, specifically, the [O\Romannum{3}]/H$\beta$ and [N\Romannum{2}]/H$\alpha$ line ratios to select star-forming spaxels \citep{2006Kewley(b)}:
\begin{equation}\label{eq:kewley_sf}
    \log\left( \frac{\mathrm{O\Romannum{3}}}{\mathrm{H\beta}} \right) < \frac{0.61}{\log(\mathrm{N\Romannum{2}}/\mathrm{H\alpha}) - 0.05} + 1.3,
\end{equation}
and require the S/N of all lines to be larger than 3.

In addition, a minimum of 10 adjacent rejuvenating spaxels is required to be a solid detection of the rejuvenating region. We identify 173 galaxies that contain at least 1 candidate rejuvenation region. 

Next, we visually inspect the spectra of these candidate rejuvenation regions to exclude interlopers. We identify 23 regions whose spectra are dominated by QSOs and broadline AGNs, which lead to incorrect spectral model in the DAP; 12 regions that are contaminated by foreground stars but not masked out; 5 M31 pointings that are parts of M31; 3 globular clusters; 3 galaxy pairs with different redshifts; and 1 Type \Romannum{2} supernova \citep{SN2019gmh}. We further exclude 14 galaxies without measurements of stellar mass and morphology and 2 galaxies for which the IFU pointings are off-center. Eventually, we have 110 galaxies containing rejuvenating regions, which are mostly localized and patchily distributed within the galaxies. In this paper, we refer them as rejuvenation hosts. 


\subsection{The Star-formation Histories (SFHs) of the Rejuvenating Regions} \label{sec:visible_time}
Here, we build SFHs combined with stellar population synthesis models to gain knowledge about the SFHs of rejuvenating regions selected by Equation~\ref{eq:Seletion_criteria}. Our model SFH is composed of two parts. The first component is a simple stellar population (SSP), representing the pre-existing old stars. The second component is the rejuvenation event, represented by a period of constant star formation, which starts $\Delta T$ years after the SSP and forms $f$ times of the stellar mass in the SSP. We examine a grid of SFHs with $1~\mbox{Gyr} \leq \Delta T \leq 10~\mbox{Gyr}$ and $ 0.2\% < f < 40\%$. 
 
We use \textsc{Flexible Stellar Population Synthesis} for python \citep[pyFSPS;][]{2009Conroy, 2010Conroy, 2024pyFSPS}\footnote{https://zenodo.org/records/12447779} to generate the mock spectra. We set a dust attenuation $A_V=0$ and stellar metallicity $Z=Z_\odot$ for the old stellar population, based on the typical stellar metallicity of quiescent galaxies with $M_* \sim 10^{11}\ \mathrm{M_\odot}$. We set $A_V=1$ for the new star-forming event \citep{2012Sobral} and allow different metallicities of $\log(Z/\mathrm{Z_\odot}) = [-0.2, 0, 0.2]$. We then measure the evolution \hd and D$_n$4000 from the onset of rejuvenation. From the evolutionary tracks, we can obtain the timescales that a region would be selected by Equation~\ref{eq:Seletion_criteria}.

\begin{figure*}[htbp]
    \centering
    \includegraphics[width=1\linewidth]{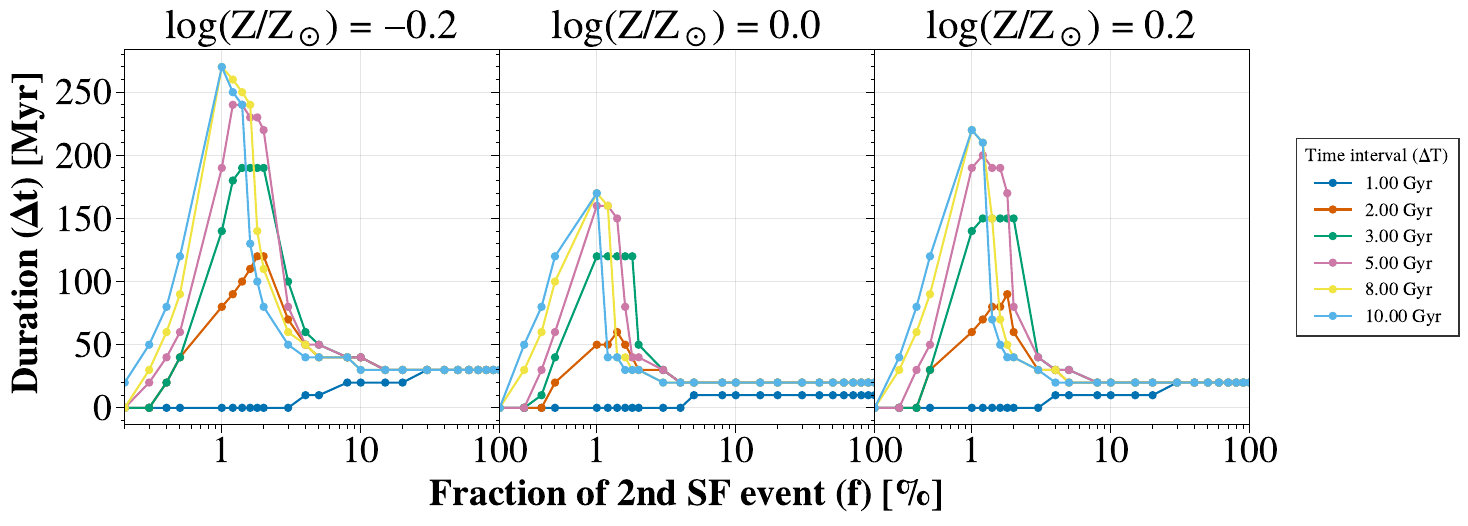}
    \caption{The visibility time ($\Delta t$) of the rejuvenation event based on our criteria (Equation \ref{eq:Seletion_criteria}). Each panel plots different stellar metallicities of the rejuvenating stellar population. Colored lines indicate different intervals between the star-formation events ($\Delta T$). Our criteria tend to select rejuvenation events that happen in the past $\lesssim 200$~Myr and have formed $\sim1\%$ of masses on top of a stellar population that is at least a few Gyr old.}
    \label{fig:falling_time}
\end{figure*}

Figure \ref{fig:falling_time} shows the visibility of rejuvenation periods at a given $\Delta T$ and $f$. A SFH with $f \sim 1\%$ would appear as a rejuvenating region for the longest period of time of $\lesssim 200$~Myrs. For slightly larger or smaller $f$, the visible period quickly drops below 100~Myr regardless of $\Delta T$. 

We then ask whether the sSFRs of these model rejuvenating regions are comparable to normal star-forming regions. We find that the sSFR of the model rejuvenating regions are mostly between $-10.4$ and $-9.72$~Gyr$^{-1}$ (the 16th and 84th percentiles), with a tail toward high sSFR of $\sim-9$~Gyr$^{-1}$. The distribution of sSFR is in broad agreement with typical star-forming regions. 

From these mock spectra, we infer that the rejuvenation regions we selected are truly close to the onset of the rejuvenation event, which is beneficial for us to investigate the properties before they are smeared out or faded away. However, our selection cannot identify strong rejuvenation events ($f \gtrsim 10\%$). Because of the low mass-to-light ratios of young stellar populations, the spectra of this kind of SFHs become similar to those of typical star-forming galaxies and thus indistinguishable by \hd and D$_n$4000. Nevertheless, those strong rejuvenation events should also have high sSFRs that can be otherwise easily identified.

\subsection{The Control Samples}
\begin{figure}
    \centering
    \includegraphics[width=\linewidth]{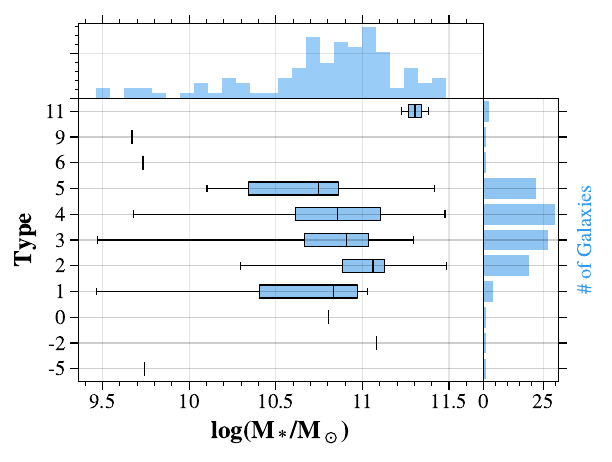}
    \caption{Distribution of stellar masses across different numerical morphological types. The associating distributions of stellar mass and morphology are shown as probability density distribution in upper sub-panel and bar chart in right sub-panel. Main panel consists of box plots where, for each type, the blue box spans the 25th percentile to the 75th percentile of the stellar mass distribution, with the vertical line inside the box indicating the median. Left and right bars represent the minimum and maximum value of stellar mass in each morphological type. Our rejuvenation hosts are generally massive and mostly late-type.}
    \label{fig:morphology_mass}
\end{figure}

We select a star-forming and a quiescent sample as control samples to be compared with rejuvenation hosts. Figure~\ref{fig:morphology_mass} shows the distribution of stellar masses and numerical morphological types of rejuvenation hosts. We find that our rejuvenation hosts are generally more massive and mostly late-types. We first require our control samples to be late-types with numerical types of $1 \leq T \leq 7$, which corresponds to from Sa to Sd, and to be face-on ($b/a>0.7$) to minimize projection effects. 

The star-forming and quiescent samples are identified by dust-corrected $\mathrm{NUV-r}$: $\mathrm{NUV-r<4}$ and $>5$ are star-forming and quiescent, respectively. We then construct mass-matched samples. For each rejuvenation host, we select all star-forming and quiescent galaxies with similar stellar masses ($\Delta \log(M_*/\mathrm{M_{\odot}})<0.2$). Our final star-forming and quiescent control samples consist of 1633 star-forming and 137 quiescent galaxies, respectively.

\section{Physical Properties of Rejuvenating Galaxies} \label{sec:local_RJGs}

In this section, we first investigate whether rejuvenation is caused by in-falling gas from the outside galaxies by examining local gas metallicities and velocities. We then present the physical properties of rejuvenation hosts to find out what kind of galaxies that rejuvenation is likely to happen and investigate the possible physical mechanisms. 

\subsection{Properties of the Rejuvenating Region} \label{sec:local_rejuvenation}

\subsubsection{Gas-phase Metallicity} \label{sec:local_metallicity}

To derive gas-phase metallicity, we adopt the empirical calibration of \cite{2013Marino}, which uses the O3N2 index, 

\begin{equation}\label{eq:gas-phase_Z}
	\mathrm{12+\log(O/H) = 8.533-0.214\times O3N2},
\end{equation}
where
\begin{equation}\label{eq:O3N2}
	\mathrm{O3N2} = \mathrm{\log\left( \frac{[O\Romannum{3}]5007}{[H\beta]4862} \times \frac{[H\alpha]6564}{[N\Romannum{2}]6583} \right)}.
\end{equation}

\begin{figure*}
	\centering
	\begin{minipage}{0.49\linewidth}
		\centering
		\includegraphics[width=\linewidth]{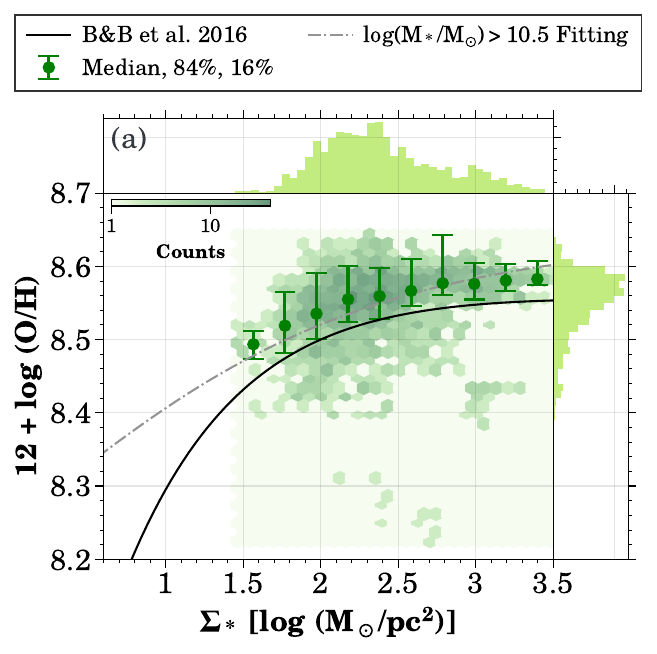}
	\end{minipage}
	\hfill
	\begin{minipage}{0.49\linewidth}
		\centering
		\includegraphics[width=\linewidth]{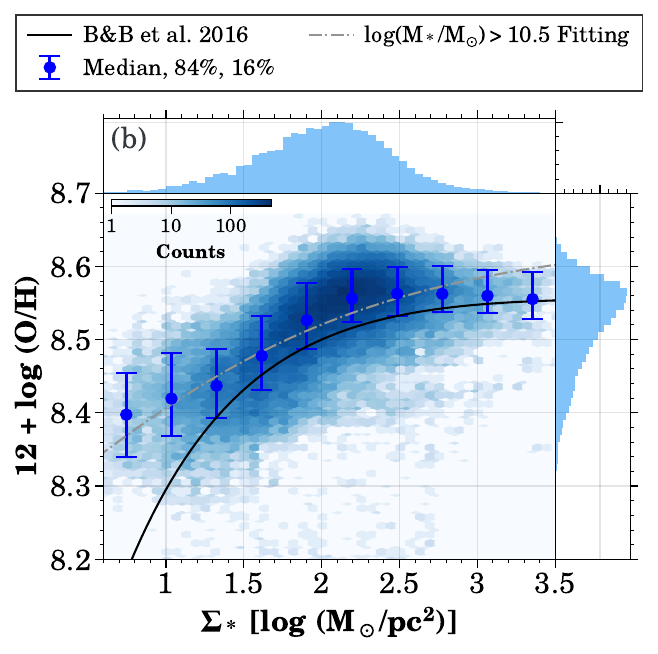}
	\end{minipage}
	\caption{The stellar mass density-metallicity ($\Sigma_*$-Z) relation for (a) rejuvenating spaxels and (b) star-forming spaxels, visualized via hexagonal binning. Each panel includes a color bar indicating the number of data points per hexagonal bin. Error bars represent the median metallicity and the 16th and 84th percentiles for each bin of $\Sigma_*$. The black solid line represents the relation derived from 507,000 star-forming spaxels in 653 star-forming disk galaxies from \citet{2016B&B}, while the gray dashed line corresponds to massive ($M_*>\mathrm{10^{10.5} \ M_\odot}$) galaxies.}
	\label{fig:MZ_relation}
\end{figure*}

Figure~\ref{fig:MZ_relation} shows the surface stellar mass density ($\Sigma_*$) and the gas metallicity of rejuvenating and star-forming spaxels in rejuvenation hosts. At a given $\Sigma_*$, the gas metallicities in rejuvenation hosts are systematically higher than the average $\Sigma_*$-Z relation (black solid line) measured from 643 star-forming disk galaxies in MaNGA \citep{2016B&B}. 

However, since the $\Sigma_*$-Z relation is stellar mass-dependent \citep{2016B&B}, we need to compare our rejuvenation hosts to galaxies of similar stellar mass. The gray dashed lines in Figure~\ref{fig:MZ_relation} are the $\Sigma_*$-Z relation derived from the $M_*>10^{10.5} \ \mathrm{M_\odot}$ SF comparison sample:
\begin{equation}
	y=8.63-1.417(x+1.97) e^{-(x-1.97)},
\end{equation}
where $x=\log(\Sigma_*[\text{M}_*/\text{pc}^2])$ and $y=12+\log(\text{O}/\text{H})$. We find that the gas metallicity in rejuvenation hosts at fixed $\Sigma_*$ is not systematically lower than it in star forming galaxies of similar stellar masses (Figure \ref{fig:MZ_relation}a). We also examine the gas metallicity of star-forming spaxels that are not classified as rejuvenating in rejuvenation hosts (Figure~\ref{fig:MZ_relation}b); similarly, the gas metallicity in rejuvenation hosts is in general not lower than normal star-forming galaxies.

While abnormally low-metallicity regions have been detected in MaNGA observations and argued to be delivered from the halo and trigger star formation \citep{2019Hwang}, we do not find such activity to be prevalent in rejuvenation hosts. Nevertheless, there is still a small fraction of spaxels that deviate significantly from the average $\Sigma_*$-Z relation. We visually inspect the spectra and confirm that the measurements are precise. We will present such a case in Section~\ref{sec:special}.

\subsubsection{Gas Velocity} \label{sec:local_gas}
We compare the gas velocities of the rejuvenating regions to that of their surroundings to search for signs of accretion. For each rejuvenating region, we select a ring of a width of 3 pixels surrounding the rejuvenating region, for which the inner boundary has a 3-pixel gap from the rejuvenating region. We then measure the median velocities from the H$\alpha$ emission in the ring and the rejuvenating region. 

\begin{figure}
	\centering
	\includegraphics[width=1\linewidth]{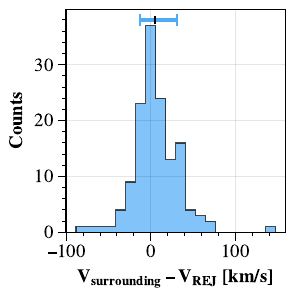}
	\caption{The distribution of the difference of gas velocities between the rejuvenating regions and their surroundings. The error bar on the top represents the 16th, 50th, and 84th percentiles of the distribution.}
	\label{fig:vel_diff_hist}
\end{figure}

Figure~\ref{fig:vel_diff_hist} shows the distribution of the differences in the gas velocities of rejuvenating regions and their surroundings. The 16th, 50th, and 84th percentiles are $-13$, 4.7, and 31 km/s, respectively. Only 2 regions deviate more than 100 km/s from their surroundings. We do not find a systematic shift in the gas velocities of the rejuvenating regions.

\subsection{Properties of the Host Galaxies} \label{sec:properties_host}

\subsubsection{Gas Metallicity Gradients} \label{sec:MG}
Radial transport of gas within a galaxy has the potential to induce star formation \citep{2012Torrey, 2019Moreno}. If gas flows from the galaxy's outskirts toward its center, the metal-poor gas originating from the outskirts will dilute the metal-rich gas in the inner region, consequently leading to a flatter metallicity gradient across the galaxy \citep{2010Rupke_a, 2010Rupke_b, 2010Kewley, 2019Moreno, 2018Bustamante}. Therefore, we compare the metallicity gradients of rejuvenation hosts to those of controlled SFGs to seek signs for radial transportation of gas. 

The gas metallicity gradients of galaxies are estimated between 0.5 and 2 effective radius ($R_e$). We measure the median metallicity in each radial bin of 0.1$R_e$ and perform an unweighted least-squares linear fit to obtain the gradients. Some galaxies do not have measurements over the full range of radii, we calculate their gradients from radii with available data. If a galaxy has less than 5 bins with valid data, we do not calculate the gradient.

\begin{figure}
    \centering
    \includegraphics[width=\linewidth]{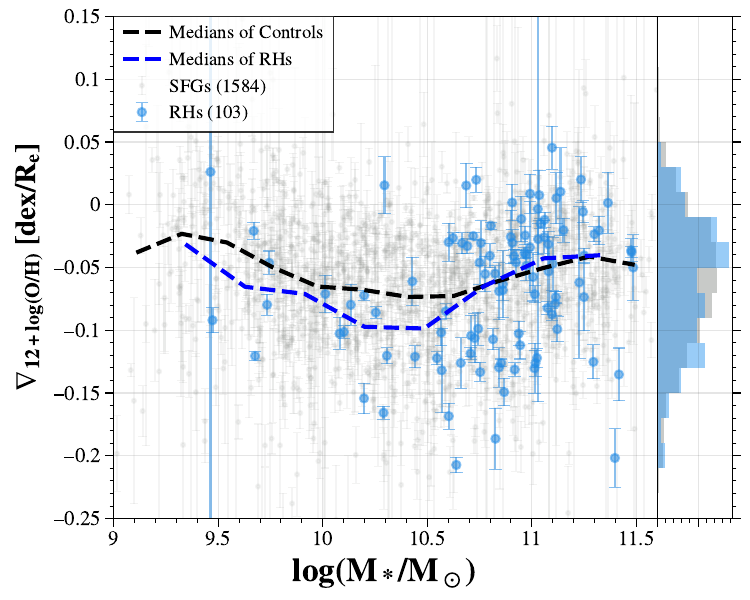}
    \caption{The gas metallicity gradients of rejuvenation hosts (blue) and SFGs (gray). The errorbars of metallicity gradients are estimated by bootstrapping. Blue and black dashed line represent the medians of gas metallicity gradients in each stellar mass bin. The side-panel represents the probability density distributions of the metallicity gradients. The metallicity gradients of rejuvenation hosts are not flatter than those of SFGs.}
    \label{fig:mg}
\end{figure}

Figure \ref{fig:mg} shows the metallicity gradients in an unit of $R_e^{-1}$ of rejuvenation hosts (blue) and SFGs (gray). The errors of the gradients are estimated by bootstrapping. We find that the rejuvenation hosts and SFGs have similar distributions of metallicity gradients. There is thus no clear supporting evidence for such gas radial transport within the rejuvenation hosts.

\subsubsection{Environmental Dependence} \label{sec:env}
Gravitational interactions with neighbors may trigger star formation within galaxies \citep{1998Kennicutt, 2007Woods, 2008Ellison, 2019Pearson, 2019Pan}. Therefore, we investigate whether the rejuvenation is related to the environments they reside in.

We compare the tidal parameters ($Q$) and the 5th-nearest-neighbor projected densities ($\eta_5$) (see Section~\ref{sec:data}) of rejuvenation hosts to SFGs and QGs. To control the mass-dependence, for each rejuvenation host, we select SFGs and QGs of similar masses ($\Delta \log(M_*/\mathrm{M_\odot}) < 0.2$) and calculate the distributions of $\Delta Q = Q_{\mathrm{RH}} - Q_{\mathrm{control}}$ and $\Delta \eta_5 = \eta_{5,\mathrm{RH}} - \eta_{5,\mathrm{control}}$. We then measure the medians of the distributions and their uncertainties by bootstrapping.

\begin{figure*}
	\includegraphics[width=0.95\textwidth]{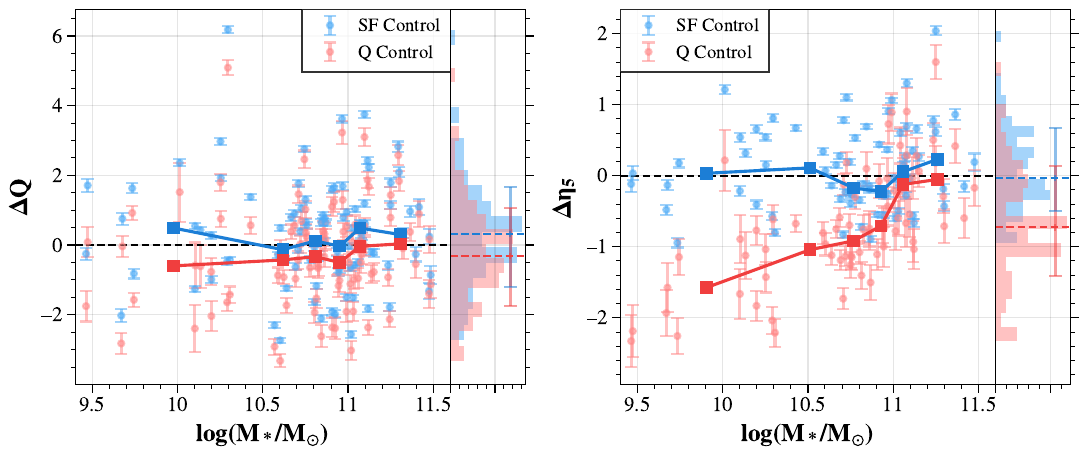}
	\caption{The distribution of $\Delta Q =Q_{\mathrm{RH}} - Q_{\mathrm{control}}$ (left panel) and $\Delta \eta_5=\eta_{5,\mathrm{RH}} - \eta_{5,\mathrm{control}}$ (right panel) of the rejuvenation hosts. Comparisons of star-forming and quiescent galaxies are shown in blue and red, and the corresponding uncertainties are estimated by bootstrapping. Squares indicate medians of $\Delta Q$ and $\Delta \eta_5$ in each equal-number bin. The probability density distributions are shown in the right sub-panels, with the 50th percentile indicated by the dashed lines and errorbars representing the 16th and 84th percentiles.}
    \label{fig:env}
\end{figure*}

Figure~\ref{fig:env} shows the $\Delta Q$ and $\Delta \eta$ for the comparison of SFGs and QGs, respectively. The distributions of $\Delta Q$ for SFGs and QGs center around zero and there is no clear mass dependence. On the other hand, the distribution of $\Delta \eta_5$ for QGs is mostly negative and shows a mass dependence, where low-mass rejuvenation hosts are located at lower-density environments than QGs of similar stellar masses. 

Overall, the environments of rejuvenation hosts are indistinguishable from SFGs. In terms of local densities, lower-mass rejuvenation hosts and SFGs live in lower-density environments. In the next section, we present our visual inspection to identify possible interaction events based on morphology.

\subsubsection{Interaction Classes} \label{sec:interaction_classes}

For the visual classification of possible post-mergers or close companions, we refer to the definition and method of the interaction stages employed in \cite{2019Pan}. They identify spectroscopic close companions of 4690 MaNGA galaxies within a projected distance of $<71.4~\text{kpc}$ and a line-of-sight velocity difference of $<500~\text{km/s}$. The five stages are:
\begin{itemize}
    \setlength\itemsep{-0.1em}
    \item Stage 0: non-interacting galaxies.
    \item Stage 1: well-separated pairs without sign of distortion.
    \item Stage 2: well-separated pairs with strong sign of interaction, such as tail and bridges.
    \item Stage 3: well-separated pairs with weak morphology distortion.
    \item Stage 4: two components are overlapping or single component with obvious tidal features, such as shells or tails.
\end{itemize}
We define Stage 0 as non-interacting, Stage 1-3 as interacting, and Stage 4 as post-merger.

\begin{figure*}
    \centering
    \includegraphics[width=1\linewidth]{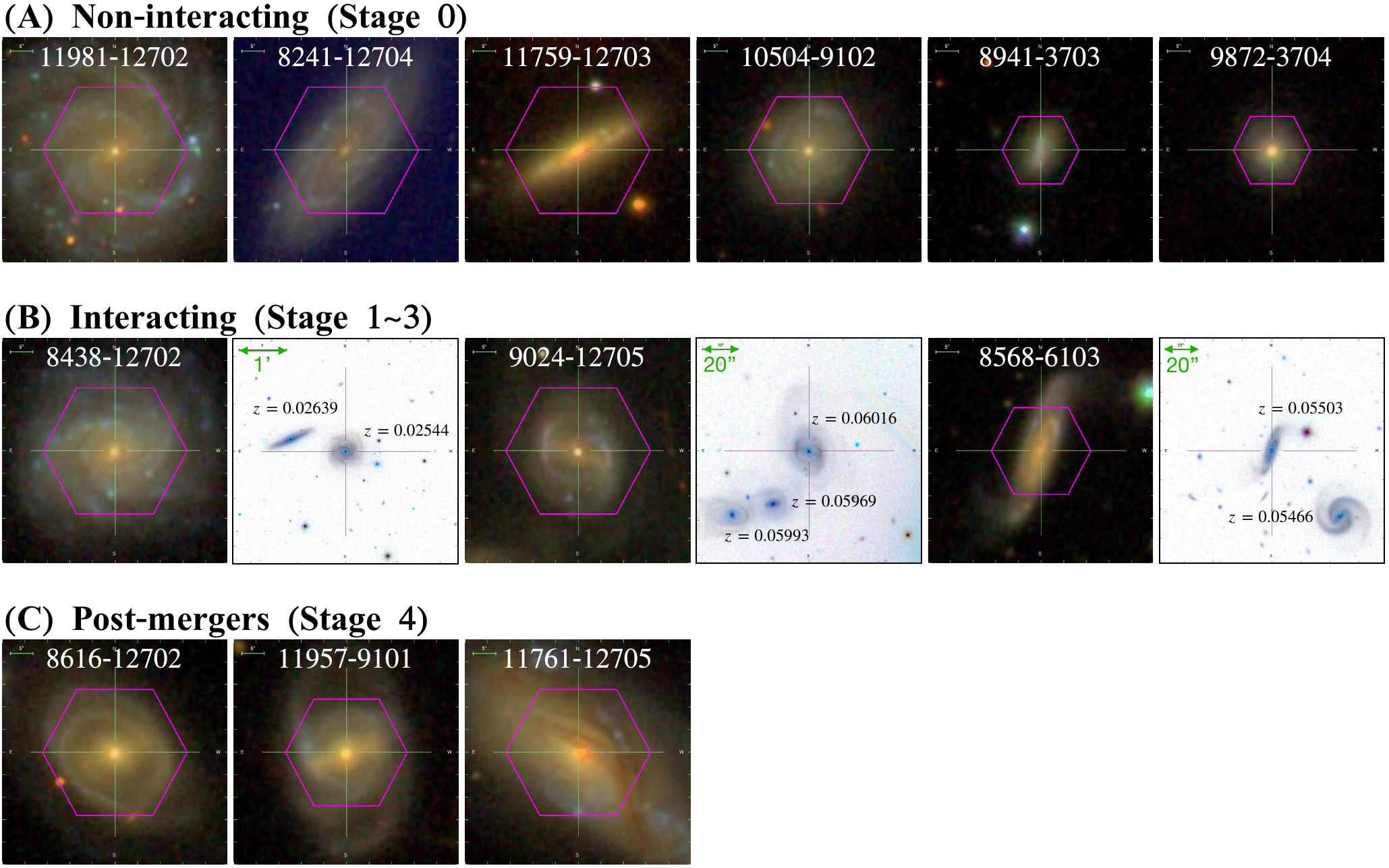}
    \caption{Example rejuvenation hosts classified as (A) non-interacting (Stage 0), (B) interacting (Stage 1--3), and (C) post-mergers (Stage 4). For Stage 1, 2, and 3, we also show a zoomed-out image to include the companion galaxies and their spectroscopic redshifts. Non-interacting galaxies are mostly late-type galaxies. Some of them show obvious spiral arms or bars. A variety of tidal features are shown in (C), including rings (MaNGA 8616-12702), tidal tails (MaNGA 11957-9101), and strong tidal disruption (MaNGA 11761-12705).}
    \label{fig:interaction_stage}
\end{figure*}

We use SDSS \textit{gri} images to perform the classification. Figure~\ref{fig:interaction_stage} shows the representative galaxies in each Stage. We find rejuvenation hosts show a variety of late-type morphologies: spiral arms, bars, and featureless disks, et cetera. For Stage 1, 2, and 3, we also show images with a wider field of view to include the companions. Overall, we find that among our 110 rejuvenation hosts, 82 are non-interacting, 21 are interacting, and 7 are post-mergers. In total, 25\% (28/110) of rejuvenation hosts are interacting or post-merger systems, comparable to the 23\% found by \citet{2019Pan} from more than 4,000 MaNGA galaxies with the same image data.

We further investigate deeper imaging and spectroscopic observations from the DESI Legacy Imaging Surveys DR9 \citep{2019Dey} and DESI DR1 \citep{2026DESI} to identify companions and tidal features that may be missed by SDSS. We find extra 4 interacting systems and no additional post-mergers, which do not significantly change the fraction of interacting rejuvenation hosts. With these 4 additional interacting systems, the fraction increases slightly to 29\% (32/110), which is still not significantly higher than that reported by \citet{2019Pan}.

\subsubsection{\ion{H}{1} Fraction} \label{sec:gas}
By definition, the progenitors of rejuvenation hosts are QGs. Therefore, we compare whether our rejuvenation hosts contain more gas than other QGs.

\begin{figure*}
    \centering
    \includegraphics[width=1\linewidth]{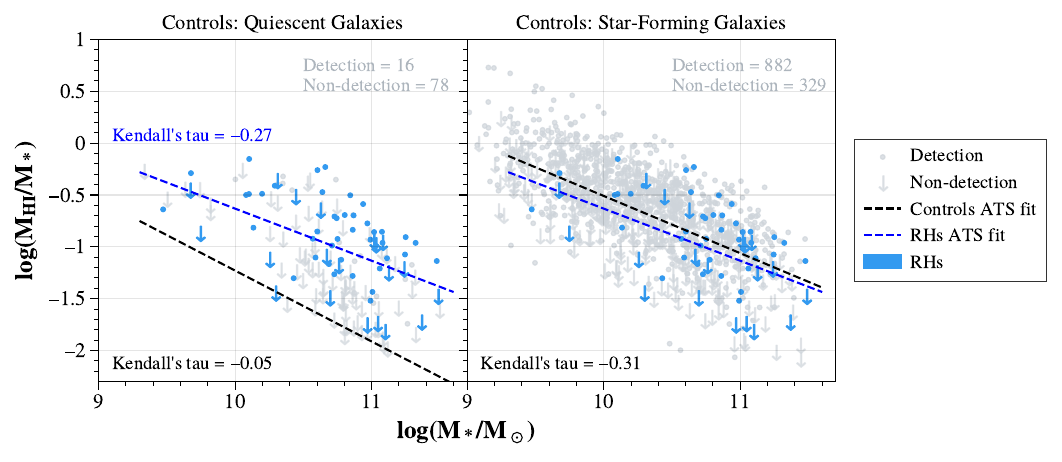}
    \caption{Stellar mass versus \ion{H}{1} fraction for rejuvenation hosts (blue) and control samples (gray). Dots and down-arrows represent detection and non-detection. The controlled samples in the left and right panels are QGs and SFGs. Black and blue dashed line shows the ATS fits for control samples and rejuvenation hosts, together with their Kendall's $\tau$s. At similar stellar masses, rejuvenation hosts have similar gas fractions to SFGs and higher gas fractions than QGs.}
    \label{fig:gas_comparison}
\end{figure*}

Figure~\ref{fig:gas_comparison} plots the \ion{H}{1} gas fraction $f_{\text{H\Romannum{1}}}=M_\mathrm{H\Romannum{1}}/M_*$ as a function of stellar mass for rejuvenation hosts (blue) and comparison samples (gray) for which with \ion{H}{1} measurements. Detections and non-detections of control samples are represented by dots and down arrows. For non-detections, we plot the 1$\sigma$ upper limits.

Within our star-forming and quiescent control samples, 1211 galaxies (comprising 882 detections and 329 non-detections) and 94 galaxies (including 16 detections and 78 non-detections), respectively, overlap with the \ion{H}{1} MaNGA DR3 galaxies. In our RJG sample, a total of 62 galaxies are observed, of which 42 are detections and 20 are non-detections.

We estimate the correlation between $f_\text{H\Romannum{1}}$ and stellar mass by performing a linear regression. To take the upper limits into account, we perform survival analysis \citep[Section 2.7]{2021Stark}. We use the \texttt{ATS} routine in  the \texttt{NADA2} \citep{NADA2} package to estimate the correlation between $f_\text{H\Romannum{1}}$ and stellar mass. The \texttt{ATS} routine is based on the Akritas-Theil-Sen (ATS) estimator \citep{1995ATS}, a nonparametric method that extends Theil-Sen estimator to estimate the slope, intercept, and Kendall's $\tau$ in a linear regression when censored data (i.e., upper limits) points are present. The Kendall's $\tau$ is a correlation coefficient that is used to quantify the statistical dependence between two variables. The $\tau$ lies within the interval $(-1,1)$: $\tau=0$ means no statistical correlation, while $\tau=1, -1$ means perfect correlation and anti-correlation.

Figure~\ref{fig:gas_comparison} shows that the \ion{H}{1} fractions of rejuvenation hosts are significantly higher than the ATS fit of QGs, and are comparable to that of SFGs of the same stellar masses. rejuvenation hosts are intrinsically gas-rich systems. 

\section{A Galaxy in Rejuvenation} \label{sec:special}

In Section~3, we show that, on average, the gas metallicities of the rejuvenating regions and star-forming regions in rejuvenation hosts are similar to those in normal SFGs. In addition, the gas velocities of rejuvenation regions are similar to their surrounding areas. 

However, in Figure~\ref{fig:MZ_relation}, there is a small fraction of spaxels showing lower gas metallicities, deviating from the average $\Sigma_*$-Z relation. These low-metallicity regions may indicate that gas infall is responsible for the rejuvenation in at least a minor fraction of galaxies. We thus further inspect galaxies that host low-metallicity regions. We find an unambiguous case that the rejuvenation of a massive galaxy triggered by an infalling gaseous component caught in the act.

\subsection{A strong emission-line region in an old galaxy}

\begin{figure*}
	\centering
	\includegraphics[width=\textwidth,height=0.85\textheight,keepaspectratio]{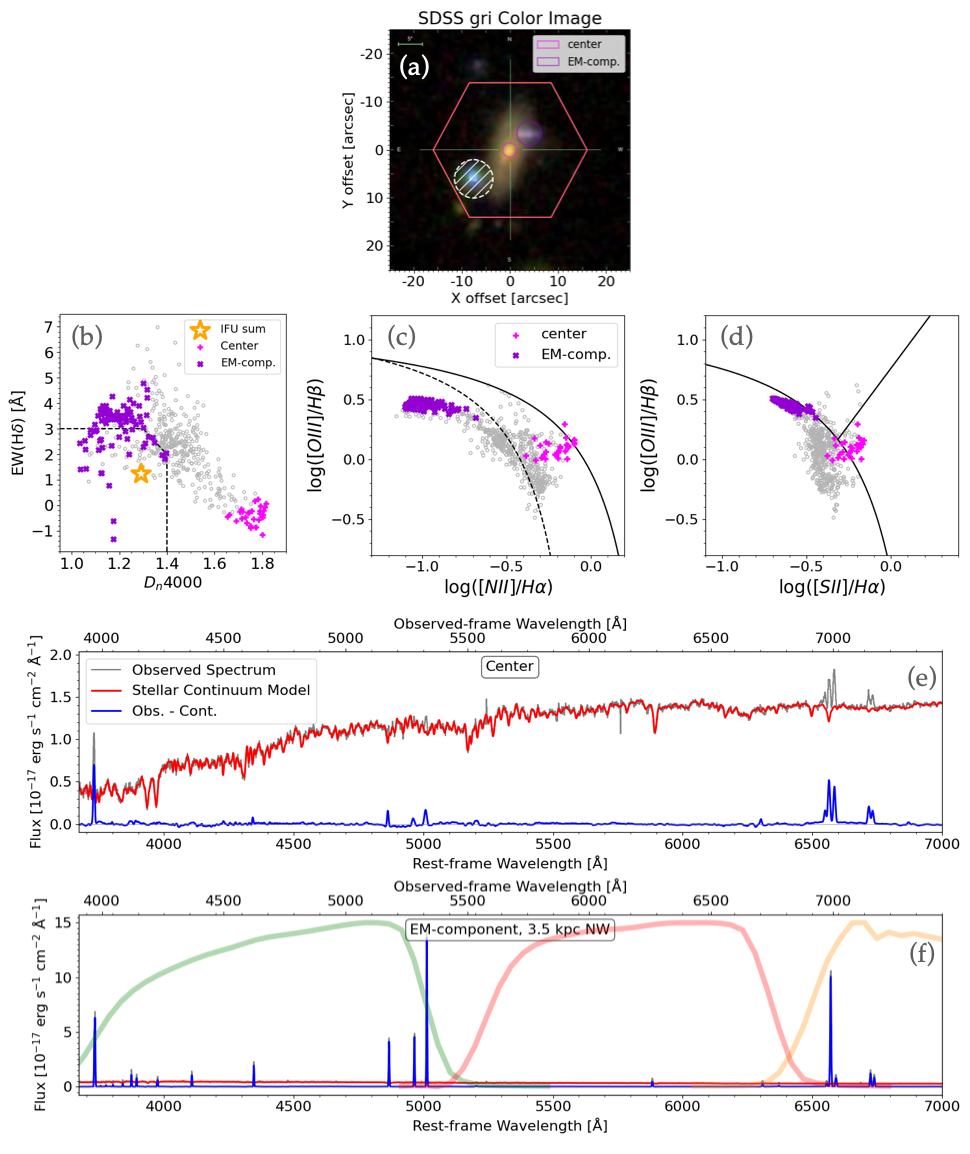}
	\caption{(a) The \textit{gri} color composite image of MaNGA 12080-12705. A purple region is located north-west of the galaxy center (purple circle). (b) The \hd and D$_n$4000 of each spaxels in the galaxy. Magenta data points are spaxels at central 1.5", corresponding to the size of an SDSS fiber. The stellar population at the center is the oldest in the galaxy. Purple data points are spaxels within a 4" circular aperture from the center of the emission-line component, where the youngest stellar population is located. This galaxy will be classified as a rejuvenating galaxy base on the spectrum integrated over the entire IFU (orange star). (c,d) BPT diagrams. The line emission at the galaxy center is LINER-like, indicating an old stellar population; whereas the emission-line region exhibits emission from low-metallicity \ion{H}{2} regions. (e,f) Spectra of the galaxy center and the center of the emission-line component. The thick green, red, and orange curves are the filter responses of SDSS $g$, $r$, and $i$ filter, respectively. The strong [\ion{O}{2}] and [\ion{O}{3}] emission in the $g$-band and H$\alpha$ emission in the $r$-band make the purple color in the composite image. \label{fig:12080-12705}}
\end{figure*}

Figure~\ref{fig:12080-12705}a shows the $gri$ composite image of MaNGA 12080-12705 with a prominent bulge. An intriguing purple region is located $\sim3.5$~kpc north-west of the center.

We find that MaNGA 12080-12705 is separated into two photometric components in the SDSS database, corresponding to the massive quiescent galaxy (hereafter, ``main component") and the purple region (hereafter, ``emission-line component"). Both components are targeted by SDSS fibers. 

Figure~\ref{fig:12080-12705}e shows the spectrum of the central spaxel of MaNGA 12080-12705. The strong D$_n$4000 break and metal absorption lines indicate an old stellar population. The central region has the oldest stellar population in the galaxy according to the D$_n$4000 and \hd (Figure~\ref{fig:12080-12705}b). The weak Balmer emission and high [S\Romannum{2}]/H$\alpha$ line ratio suggest a LINER-like emission (Figure~\ref{fig:12080-12705}d), which is also a signature of an old stellar population.

On the contrary, the spectrum at the location of the emission-line component shows prominent emission lines (Figure~\ref{fig:12080-12705}f). The purple color in the $gri$ composite color image is driven by the strong [O\Romannum{2}] and [O\Romannum{3}] emissions in the $g$-band and H$\alpha$ emission in the $i$-band, which boost up the total fluxes in the red and blue channels. We show the emission line ratios of spaxels within 4", roughly 2$R_e$ of the emission-line region in the $u$-band, from the center of the emission-line region (the purple circle in Figure~\ref{fig:12080-12705}a). The low [N\Romannum{2}]/H$\alpha$ and [S\Romannum{2}]/H$\alpha$ line ratios indicate low-metallicity H\Romannum{2} regions as the sources of ionizing photons (Figure~\ref{fig:12080-12705}c,d). 

\subsection{Stellar Mass and Star-formation Rate}

We take the \texttt{modelMag} in 5 SDSS bands and the spectroscopic redshift for each component, and use \texttt{kcorrect} version 5.0.0 \citep{2007Blanton} to derive the stellar mass of $10^{10.85} \ \mathrm{M_\odot}$ and $10^{9.0} \ \mathrm{M_\odot}$ for the main component and the emission-line component, respectively. The stellar mass of the main component we derived is consistent with the stellar masses categorized in the NASA Sloan Atlas (NSA) at $\sim0.1$~dex level.

We then estimate the SFR of the emission-line component from its H$\alpha$ luminosity and correct for the dust attenuation using the Balmer decrement, following the prescription in \citet{2004Brinchmann}. It is non-trivial to completely separate the line emission from the H\Romannum{2} region and LINER-like emission from old stars, therefore, we estimate two SFRs from two different apertures. First, we use a circular aperture centered at the emission-line component with a radius of 4", i.e., the purple circle in Figure~\ref{fig:12080-12705}a. The second estimate uses the line fluxes integrating over the entire galaxy \citep{2021Wu}. We take these two values as the lower and upper limits of our SFR estimates. 

We thus obtain an $\mathrm{SFR}_{\mathrm{H\alpha}} = 2.7 - 3.6\ \mathrm{M_\odot}$~yr$^{-1}$. Compared to the star-formation main sequence obtained by the same prescription for stellar mass and SFR, $\log(\mathrm{SFR} / [\mathrm{M_\odot} \ \mbox{yr}^{-1}]) = 0.76 \times \log(M_* / \mathrm{M_\odot}) - 7.64$ \citep{2015Renzini} by assuming $M_*=10^{9.0} \ \mathrm{M_\odot}$, the SFR of the emission-line component is $\sim 20$ times higher than typical SFGs of similar stellar masses. The stellar mass in the emission-line component can form within a short time, less than 300 Myrs, if the current SFR has been sustained in the past.






\subsection{Gas-phase Metallicity and Kinematics}

\begin{figure*}
	\centering
	\includegraphics[width=0.95\textwidth]{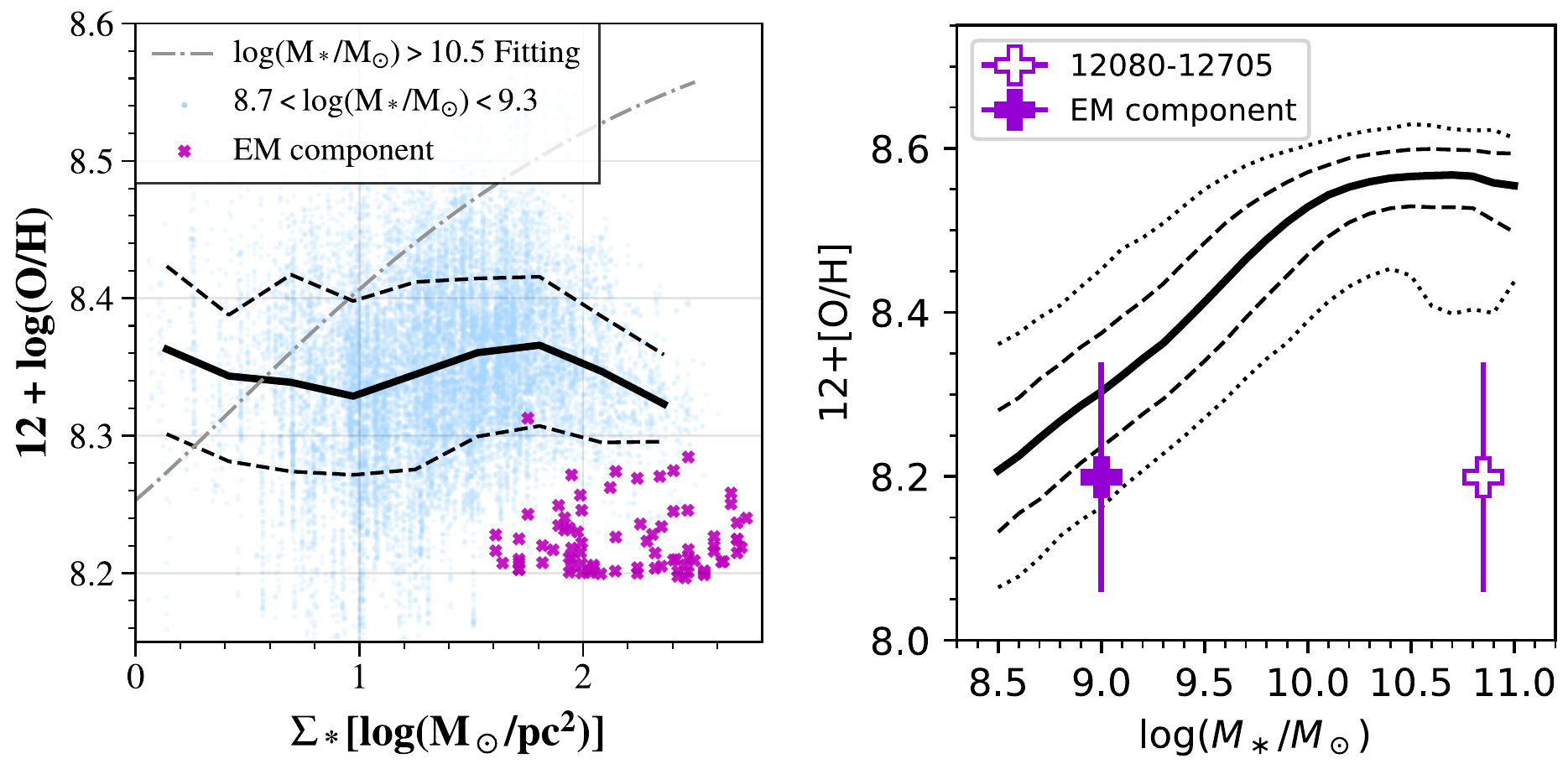}
	\caption{\textit{Left:} Comparison to $\Sigma_*$-Z relation. The blue points are BPT star-forming spaxels selected from 32 low-mass SFGs with stellar masses $8.7<\log(M_*/\mathrm{M_{\odot}})<9.3$, and the black lines represent the 16th, 50th, and 84th percentiles of metallicities for a given surface stellar mass density ($\Sigma_*$). The gray dashed line is same to that in Figure \ref{fig:MZ_relation}. The emission-line component remains extremely metal-poor compared to other star-forming regions in galaxies with similar stellar masses. \textit{Right:} Comparison to the mass-metallicity relation. The lines are the 5th, 16th, 50th, 84th, and 95th percentiles of the mass-metallicity relation derived from SDSS galaxies at the same redshift range as MaNGA galaxies. MaNGA 12080-12705 is extremely metal-poor when considering the mass of the main component (open cross). If only the stellar mass of the emission-line component is counted, the galaxy is still below the mean mass metallicity relation but less extreme (filled cross). The errorbar represents the typical uncertainty from the metallicity calibration of 0.14~dex. The uncertainty from the line fluxes are smaller than than the size of the markers.}
    \label{fig:mz}
\end{figure*}

Figure~\ref{fig:mz}(a) compares the the gas metallicities within the emission-line component to the local $\Sigma_*$-Z relation derived in Section \ref{sec:local_RJGs} (gray dashed line). The metallicities of the emission-line components are $\sim 0.3$ dex lower than the $\Sigma_*$-Z relation calculated for galaxies with $M_\ast \simeq 10^{11} \ \mathrm{M_\odot}$. Even when compared to the $\Sigma_*$-Z relation of MaNGA SFGs with $M_\ast \simeq 10^{9} \ \mathrm{M_\odot}$, the metallicities remain lower than those of typical low-mass SFGs. The low metallicity is qualitatively consistent with the anti-correlation between metallicity and SFR, which could be explained as strong gas inflow of metal-poor gas dilutes the metallicity \citep{2014Zahid,2010Mannucci}.





We then integrate the line fluxes within the emission-line component (the purple circle in Figure \ref{fig:12080-12705}a) and measure a global metallicity $12+\mathrm{[O/H]} \simeq 8.2$ from the integrated fluxes. We compare this metallicity to the global mass-metallicity (M-Z) relation of SDSS emission-line galaxies at similar redshifts using \cite{2013Marino} calibration (black lines in Figure \ref{fig:mz}(b)). We find that the global metallicity of MaNGA 12080-12705 is significantly lower than that of galaxies with $M_\ast \simeq 10^{11} \ \mathrm{M_\odot}$ (open cross). If we instead compare it to galaxies with $M_*\sim10^9 \ \mathrm{M_\odot}$ (filled cross), the metallicity of the emission-line component is still below the average, although less dramatic.

\begin{figure*}
	\centering
	\includegraphics[width=0.95\textwidth]{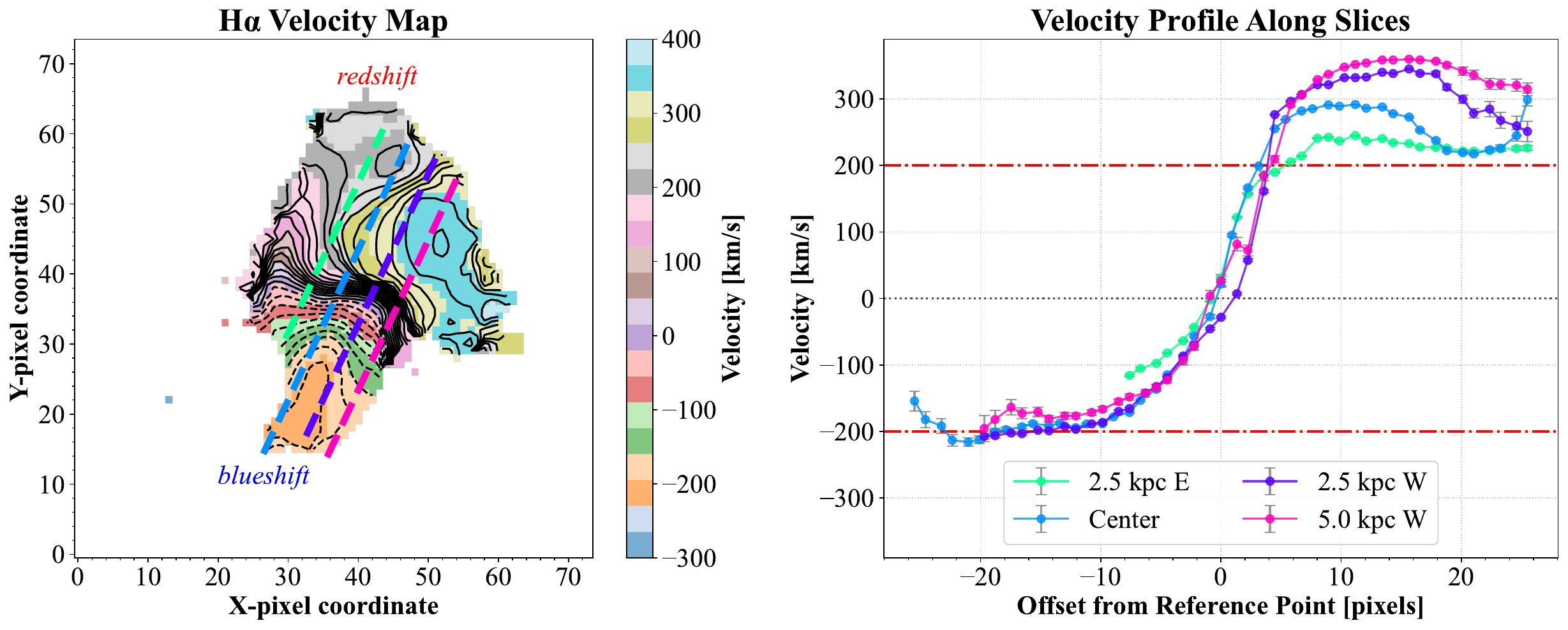}
	\caption{\textit{Left}: The H$\alpha$ velocity map of MaNGA 12080-12705, with the foreground star in the south-east being masked out. \textit{Right}: The velocity profiles along each slice on the H$\alpha$ velocity map. In the left panel, the gas velocities reach $\sim 200 $ km/s on both the redshift (gray) and blueshift sides (orange), while the emission-line component shows a kinematically distinct velocity pattern (cyan) which reaches $\sim 350$ km/s. In the right panel, all slices converge to $\sim 200$ km/s at large radii on the blueshift side. Conversely, on the redshift side, only the eastern slice (light green) flattens near 200 km/s; the other three slices exhibit a velocity bump exceeding 300 km/s at an offset of 10 pixels. These results indicate the north-west emission-line component does not follow the rotation of MaNGA 12080-12705, representing a separate kinematic component.}
	\label{fig:12080-12705_vel}
\end{figure*}

Figure \ref{fig:12080-12705_vel} presents the H$\alpha$ gas velocity of the galaxy. As shown in the left panel, MaNGA 12080-12705 generally exhibits a velocity gradient along its major axis; however, the emission-line component (colored in cyan) deviates significantly from the galaxy's primary rotation. The right panel depicts velocity profiles from the 4 slices along the major axis, passing through the galaxy center and at offsets of 2.5" to the east and west. While velocities typically approach $\sim 200$ km/s at large radii on both the blueshift and redshift sides, the western slices exhibit elevated velocities $>300$ km/s specifically on the redshift side. This result indicates that the emission-line region does not follow the rotation of the galaxy, being a distinct kinematic component.

\subsection{MaNGA 12080-12705 as a Rejuvenating Galaxy}
From the distinct kinematics and the low metallicity of the emission-line component, we conclude that MaNGA 12080-12705 is an old, massive galaxy experiencing rejuvenation triggered by infalling material.

We stack the spectra from individual spaxels over the entire IFU plate to mimic the kind of data obtained by fiber or slit spectroscopy for distant galaxies \citep{2021Wu}. We measure $\mathrm{D}_n4000=1.29$ and $\mathrm{EW(H\delta_A)}=1.24~\text{\AA}$, which place MaNGA 12080-12705 in the rejuvenation corner (Figure~\ref{fig:12080-12705}b). The integrated stellar mass ($10^{10.85} \ \mathrm{M_\odot}$) and SFR ($\sim 3 \ \mathrm{M_\odot}\ \mbox{yr}^{-1}$) put MaNGA 12080-12705 well in the star-forming main sequence \citep{2015Renzini}.  

If MaNGA 12080-12705 were at a larger cosmological distance, our criteria allow us to identify its current rejuvenation event and distinguish it from other normal star-forming galaxies. Even though this kind of rejuvenation event is rare, it should be still possible to be identified in existing and future large scale deep spectroscopic surveys across redshifts, and would help us understand the possible underlying mechanisms responsible for the rejuvenation (see discussion in Section~\ref{sec:discussion}).

\section{Discussion} \label{sec:discussion}

\subsection{The Origin of Local Rejuvenation} \label{sec:origin_of_gas}

We investigate the physical properties of both the local rejuvenating regions and their host galaxies to constrain the underlying mechanisms driving this phenomenon. Systematically, we find that the gas-phase metallicities of rejuvenating regions are on average not lower than those of normal star-forming regions. Furthermore, the gas velocities of these regions are comparable to their surroundings, showing no evidence of distinct velocity components. At the global scale, rejuvenation hosts are predominantly morphological late-type galaxies. They exhibit \ion{H}{1} gas masses, gas-phase metallicity gradients, and environmental distributions that are overall consistent with normal SFGs. In short, the rejuvenating regions and their hosts do not show significant differences from typical star-forming systems in any of the parameters we examined.

If rejuvenation is primarily fueled by the accretion of pristine gas from the intergalactic or circumgalactic medium (IGM/CGM), one would expect to observe distinct gas kinematics and lower gas metallicities within the rejuvenating regions. Star-forming regions with chemical dilution have been reported in observations of galaxies up to $z \sim 1$ \citep[e.g.,][]{2019Hwang, 2025Estrada-Carpenter}. Furthermore, \citet{2026Lazarus} used the ratio of $\mathrm{H}\alpha$ to near-ultraviolet (NUV) emission to identify galaxies rejuvenated within the last $\sim 10~\mathrm{Myr}$, claiming statistically lower gas metallicities in these systems. 

However, the rejuvenating regions in our sample do not exhibit an overall metallicity deficit. It is therefore natural to infer that the localized rejuvenation observed in our sample is predominantly fueled by gas reservoirs already residing within the host galaxies. In this scenario, star formation was temporarily suppressed or halted in these regions before being revived. 
The internal revival and suppression aligns with the theoretical framework in which interstellar gas undergoes continuous cycle between diffuse non-star-forming and dense star-forming states on timescales of $\sim100$ Myr.  Gas within the host galaxies is periodically driven into a dense, star-forming molecular phase primarily through compression by spiral density waves and turbulent flows. These star-forming regions are rapidly disrupted and expelled back into a non-star-forming state via stellar feedback before most gas can be consumed \citep{2017Semenov,2018Semenov}. Rejuvenation hosts may be typical star-forming galaxies and the observed localized rejuvenation represents the active phase in these continuous cycles. 

This local rejuvenation is likely mostly an internal process. Our visual classification (Section~\ref{sec:interaction_classes}) reveals that the fraction of interacting or post-merger systems among rejuvenation hosts is comparable to that of the general MaNGA population. Furthermore, rejuvenation hosts do not experience stronger tidal forces from their companions compared to control samples (Figure~\ref{fig:env}). Our visual inspection also rules out an excess of faint, low-mass dwarf companions that might have been missed in the tidal parameter calculation.

Meanwhile, our results still demonstrate that rejuvenation triggered by gas accretion from outside of the galaxy can happen, as shown by MaNGA 12080-12705 (Section~\ref{sec:special}). This kind of events is likely at most a minority among rejuvenation hosts so that the average metallicities and kinematics of rejuvenating regions stay consistent with normal star-forming regions.

However, we also need to consider the timescale on which the accreting gas mixes with the resident gas in the galaxy. If we observe the galaxy long after the gas accretion event, the distinct chemical and kinematic signatures of infalling gas will be washed out by turbulent mixing in galactic disks. Numerical simulations suggest scale-dependent mixing timescales, ranging from $\sim 30~\mathrm{Myr}$ at kiloparsec scales up to $\sim 300~\mathrm{Myr}$ at the global galactic scale \citep{2002deAvillez, 2012Yang, 2015Petit}.

If the rejuvenation is in fact triggered by gas accretion from outside the galaxy, the absence of metallicity deficits and velocity offsets in our rejuvenating regions provides constraints on these mixing timescales. Our selection method is capable of identifying rejuvenating regions up to $\sim 200~\mathrm{Myr}$ after the onset of rejuvenation. At this upper limit, detectability is restricted to a narrow range of burst strengths. For younger events ($\lesssim 100~\mathrm{Myr}$), a much wider range of burst strengths can be successfully identified (see Figure~\ref{fig:falling_time}). We can therefore reasonably estimate that the majority of the rejuvenating galaxies in our sample are observed within $\sim 100~\mathrm{Myr}$ of the onset of rejuvenation. Consequently, the mixing timescale is likely to be much shorter than $\sim 100~\mathrm{Myr}$; otherwise, we would expect to detect metallicity or velocity anomalies in a larger fraction of our sample. This empirical constraint is consistent with the lower end of the mixing timescales estimated by numerical simulations \citep{2012Yang}.

\subsection{Toward Understanding the Origin of Rejuvenation across Cosmic Time}

Previous studies using full-spectral fitting claim rejuvenation events happen in galaxies by identifying a second peak in the star formation histories \citep{2019Chauke,2024Tanaka}. The full-spectral fitting does not differentiate the mass gained from mergers or in-situ star formation. The stellar masses in the second peak can be brought by a dry merger event with a galaxy of a younger stellar population. Furthermore, these studies discuss rejuvenation events happened long time ago. There is little we can know about the mechanisms that trigger the rejuvenation from those galaxies. In principle, full-spectral fitting is able to identify galaxies with current rejuvenation events. However, in practice, quantifying the SFHs of SFGs is less reliable than for QGs \citep{2020Lower, 2025Wang}. Moreover, the qualities of spectra, such as S/N, wavelength coverage, and spectral resolution affect the precision and accuracy of the SFHs and the dependence is not understood \citep{2019Carnall}.

MaNGA 12080-12705 demonstrates an alternative and viable method to catch rejuvenation events in action. Following the theoretical work of \cite{2023Zhang}, we present that current rejuvenation can be identified by just two of the most prominent stellar absorption features, D$_n$4000 and EW(H$\mathrm{\delta_A}$). Current and future large-scale spectroscopic surveys such as WEAVE-StePS \citep{2023Iovino}, LEGA-C \citep{2021vanderWel}, PFS SSP \citep{2022Greene}, and MOONRISE \citep{2020Maiolino} will provide a vast amount of high-resolution, high-S/N, rest-frame optical spectra of galaxies from the nearby Universe to $z>2$. The proximity in wavelength of D$_n$4000 and H$\delta$ further makes them an efficient combination; they are likely both in the wavelength coverage of the same spectrograph. A large number of rejuvenating galaxies can be expected to found in these surveys for statistical studies.

Taking MaNGA 12080-12705 as an example, the intense local star formation triggered by infalling gas is easily identified by the morphology and spatially-resolved broadband SED. Beyond the nearby Universe, Euclid \citep{2025Mellier} and Roman Space Telescope \citep{2019Akeson} will provide wide-field high-resolution images in multiple wavelengths. The synergy of ground-based spectroscopic surveys and space-based imaging surveys will enable a systemic, statistic investigation of mechanisms that drive rejuvenation across cosmic time. 


\subsection{Caveats and Challenges}

Here, we point out a few caveats and challenges in our analysis. Firstly, we exclude spaxels where the spectra exhibit broad-line emission or where the AGN continua are strong. In these spaxels, the absorption lines indices measured by MaNGA DAP do not represent the underlying stellar populations and formation history. Because they are AGN-related, we always find them at the center. However, in the scenario that galaxy interaction triggers gas inflow and star formation, the strongest star formation is expected to be at the galaxy center \citep{2006Kewley(a)}. If the gas fueling AGN could at the same time trigger central rejuvenation, our analysis will not pick up this kind of rejuvenation event. In principle, simultaneously modeling the AGN and stellar continuum can help recover the underlying stellar population \citep[e.g.,][]{2021Sexton}, but the accuracy and precision depend on the fractional contribution from AGN emission. In this work, we choose to adopt the measurements from MaNGA DAP for all galaxies for internal consistency. Studies of central rejuvenation regions will need further investigation.

Secondly, our selection may not capture the most intense rejuvenation events. Our illustrative simple model SFHs show that the sSFRs of rejuvenation regions selected by our criteria are comparable to normal star-forming regions (Section~\ref{sec:data}). \citet{2023Zhang} arrived qualitatively the same conclusion after examining a more comprehensive library of SFHs. If the sSFR is high enough, the light from older stellar populations is outshined by the newly-formed stars. In the D$_n$4000-\hd parameter space, the rejuvenating regions are indistinguishable from normal star-forming regions. This issue is fundamentally rooted in the evolution of stellar population and will always be difficult to overcome for observations. However, this outshining may not be a big issue; such intense events will have sSFRs higher than the star-forming main sequence. Our method is effective to identify rejuvenation regions in the star-forming main sequence.

More \ion{H}{1} data should provide better information on rejuvenation. In this work, we investigate the global \ion{H}{1} fraction in Section \ref{sec:gas} and conclude that the rejuvenation hosts are \ion{H}{1}-rich. However, since we identify local rejuvenation events, the global \ion{H}{1} masses are not representative for the local gas properties.  Spatially-resolved \ion{H}{1} surveys, such as WALLABY \citep{2020Koribalski_WALLABY} and APERTIF \citep{2022Adams,2022Apertif}, may help us better address the spatial distribution of \ion{H}{1} gas within the rejuvenation hosts; especially the latter overlaps better with the MaNGA data as it mainly targets the northern sky.

\section{Summary and Conclusion} \label{sec:conclusion}
From the MaNGA survey, we identify 110 galaxies that host rejuvenating regions, based on two stellar absorption features, D$_n$4000 and EW(H$\delta_\mathrm{A}$). These rejuvenation regions are likely undergoing a recent secondary star formation event within the last $\sim200$~Myr. 

We analyze the gas-phase metallicities, metallicity gradients, environments, and \ion{H}{1} gas fractions of the rejuvenation hosts and compare them to controlled samples of star-forming and quiescent galaxies. In addition, we visually classify the interaction stages of rejuvenation hosts. We find the following properties of rejuvenation hosts. 

\begin{itemize}
	\setlength\itemsep{-0.1em}
    \item Gas metallicity in rejuvenating regions is overall similar to that in star-forming regions of the controlled SFGs.
	\item The metallicity gradients of the rejuvenation hosts are not shallower than those of the controlled SFGs.
	\item The gas velocities in rejuvenating regions show no significant difference from those in their surrounding areas.
	\item The rejuvenation hosts experience similar strength of tidal force from neighboring galaxies to controlled SFGs and QGs, but are located at lower-density regions than QGs when the environment is quantified as the projected 5th-nearest-neighbor density.	
	\item Rejuvenation hosts exhibit higher \ion{H}{1} fraction compared to other QGs and are comparable to SFGs.
    \item Visual classification shows 25\% (28/110) rejuvenation hosts are interacting systems or post-mergers. The fraction is similar to the underlying MaNGA sample. If considering the extra 4 interacting systems classified by the DESI DR1, the interacting fraction slightly increases to 29\% (32/110), which still shows no significant increase.
\end{itemize}

Overall, among the properties we investigate, rejuvenation regions and their hosts are similar to normal star-forming regions and star-forming galaxies. Local rejuvenation events are preferentially fueled by gas that originated within galaxies. They represent the active phases in which the gas is being compress and form dense, star-forming cloud at small scale. On the other hand, if the rejuvenation is fueled by gas accretion, our observations put empirical constraints on the chemical and dynamical mixing time scales. The accreted gas has to be mixed with resident ISM in $\lesssim 100$ Myr so that no chemical or kinematical anomaly are detected.

Meanwhile, we demonstrate a representative galaxy, MaNGA 12080-12705, where an old, massive quiescent galaxy is brought back to the star-forming main sequence by accreting gas or merging with a dwarf galaxy. If MaNGA 12080-12705 were at a large cosmological distance and observed by a fiber or slit spectrograph, the D$_n$4000 and \hd will identify it as a rejuvenating galaxy and distinguish it from other star-forming galaxies.

The simplicity of the selection method makes it easily applicable to several existing and coming large-scale spectroscopic surveys. A large number of rejuvenating events should be captured by various survey data sets across a wide range of redshifts. Combing data from ground-based spectroscopic and space-based imaging surveys, we are toward understanding the mechanisms driving galaxy rejuvenation from the nearby Universe to the cosmic noon.

\section{Acknowledgments}
The authors thank the anonymous referee for providing constructive comments that solidified the conclusion of this paper. This work is supported by the Ministry of Science and Technology of Taiwan under the grants 113-2112-M-002-027-MY2 and 111-2112-M-002-048-MY2.
\vspace{5mm}


\software{Astropy \citep{astropy:2013, astropy:2018, astropy:2022},
          Marvin \citep{Marvin},
          Numpy \citep{Numpy},
          Scipy \citep{2020SciPy},
          Proplot (\url{https://proplot.readthedocs.io/en/latest/index.html}; based on Matplotlib \citep{2007Matplotlib}),
          pyFSPS \citep{2009Conroy, 2010Conroy, 2024pyFSPS},
          NADA2 \citep{NADA2},
          kcorrect \citep{Blanton2017}.
          }

\bibliography{main}{}
\bibliographystyle{aasjournal}

\end{document}